# Integration of Communication and Computational Imaging

Zhenming Yu, *Member*, *IEEE*, Liming Cheng, Hongyu Huang, Wei Zhang, Liang Lin and Kun Xu

*Abstract*—Communication enables the expansion of human visual perception beyond the limitations of time and distance, while computational imaging overcomes the constraints of depth and breadth. Although impressive achievements have been witnessed with the two types of technologies, the occlusive information flow between the two domains is a bottleneck hindering their ulterior progression. Herein, we propose a novel framework that integrates communication and computational imaging (ICCI) to break through the inherent isolation between communication and computational imaging for remote perception. By jointly considering the sensing and transmitting of remote visual information, the ICCI framework performs a full-link information transfer optimization, aiming to minimize information loss from the generation of the information source to the execution of the final vision tasks. We conduct numerical analysis and experiments to demonstrate the ICCI framework by integrating communication systems and snapshot compressive imaging systems. Compared with straightforward combination schemes, which sequentially execute sensing and transmitting, the ICCI scheme shows greater robustness against channel noise and impairments while achieving higher data compression. Moreover, an 80 km 27-band hyperspectral video perception with a rate of 30 fps is experimentally achieved. This new ICCI remote perception paradigm offers a high-efficiency solution for various real-time computer vision tasks.

## I. INTRODUCTION

Optical information plays a crucial role in enabling humans to comprehend the external world, serving as an indispensable medium in today's information era. Since the advent of charge-coupled devices (CCD) in 1970 [1], optical information has been captured and stored in digital devices, significantly expanding people's visual perceptions of the outside world. Recently, to break the physical limitations of conventional photoelectrical imaging systems, computational imaging (CI) [2-3] has emerged, aiming at incorporating computational design into image capturing procedures to improve imaging quality [4-6], expand imaging capabilities [7-9], and acquire high-dimensional optical information [10-12]. By encoding the light field and performing computational reconstruction, CI remarkably enhances the optical information interpretation capability of imaging systems. Despite broadening the depth and breadth of visual information perception, current CI techniques are still inadequate when it comes to meeting the demand for the high-throughput and high-speed remote vision tasks that necessitate information interaction over time and distance. In practical scenarios such as autonomous driving, Internet of Things, and remote sensing, visual information transmission has become an urgent issue to be resolved. Hence, integrated communication and computational imaging will become an inevitable trend in optical sensing evolution. Over the past few decades, digital communication has constituted the backbone of all forms of information infrastructure, carrying massive data traffic in modern society. With the increased demand for high-fidelity and high-capacity image transmission, advanced source coding, channel coding, certain techniques, including modulation, multiplexing, and compensation techniques, have been introduced [13-20], dramatically improving communication efficiency. However, because of the technical gap between communication and CI, the existing communication systems will inevitably induce cross-domain impairments to the CI systems. Moreover, the straightforward combination of conventional digital communication systems and CI systems has complicated information processing procedures and slow responses, hindering its deployment into practical scenarios. Therefore, a concise and task-oriented integrated framework incorporating communication and CI is needed to meet the increasing demand for remote computer vision (CV) tasks.

Essentially, communication and CI pursue a similar goal of optimal information transfer. Both technologies employ information coding strategies. Drawing inspiration from their essence, we propose the Integration of Communication and Computational Imaging (ICCI) framework to bridge this technical gap. The novel framework simultaneously extends the time, distance, dimensions and scales of remote visual

Manuscript received × × ××; revised × × ××. This work was financially supported by the National Key R&D Program of China (No.2023YFB2905900); the National Natural Science Foundation of China (No.62371056); the Shenzhen Science and Technology Program (KJZD20230923115202006); the Fund of State Key Laboratory of Information Photonics and Optical Communication BUPT (No. IPOC2021ZT18); and the Fundamental Research Funds for the Central Universities (No.530424001). *(These authors contributed equally: Zhenming Yu and Liming Cheng)* (*Corresponding author: Zhenming Yu, Kun Xu*)

Zhenming Yu, Liming Cheng, Hongyu Huang, Wei Zhang, Liang Lin and Kun Xu are with State Key Laboratory of Information Photonics and Optical Communications, Beijing University of Posts and Telecommunications, Beijing 100876, China. (e-mails: yuzhenming@bupt.edu.cn; LMCheng@bupt.edu.cn; hongyuhuang@bupt.edu.cn; zw_bupt@bupt.edu.cn; aliang@bupt.edu.cn; xukun@bupt.edu.cn).

information perception, breaking the isolation between information source acquisition and information transmission. Taking advantage of deep learning techniques [21-25], we design an ICCI encoder–interpreter network for effective remote visual information coding and interpretation. Leveraging the end-to-end learning capabilities of deep learning, the ICCI framework performs full-link optimization and realizes optimal remote perception with a concise structure. To demonstrate the performance of the ICCI scheme, we conduct numerical analysis with the coded aperture snapshot compressive imaging systems [26, 27] under additive white Gaussian noise (AWGN) channel and slow fading channel to achieve remote hyperspectral and remote high-speed video information perception. We compare the performance of the ICCI scheme with the traditional schemes in terms of the data compression ratio and time complexity. For further validation, we physically implement and demonstrate the ICCI framework using an IM-DD optical fiber transmission system and a reflective coded aperture snapshot spectral imager (R-CASSI) [28]. Fast remote four-dimensional hyperspectral video perception is achieved with the experimental setup. The results illustrate that the ICCI framework achieves more robust transmission against channel noise and impairments, higher data compression, lower time complexity, and better remote visual information recovery compared with traditional schemes. We believe that our work can provide a novel perspective to comprehend communication and imaging and inspire more intelligent and concise end-to-end ICCI designs to be applied in military electronics, civil detection, and some emerging CV applications, i.e., Metaverse.

## II. PRINCIPLE

Fig. 1(a) depicts the remote perception pipeline realized by combining communication and CI directly. As illustrated in the left part of Fig. 1(a), CI hardware performs optical projection over the plenoptic function $X(x,y,z,\lambda,t,\theta,\varphi)$, which was first introduced by Adelson in 1991 [29]. By encoding the light field, the sensors obtain measurements of the remote optical information. The right part describes the information transmission phase. Based on digital communication systems, the measurement is separately processed by source coding, channel coding, and modulation. By applying advanced source encoding techniques, the input measurement is encoded into a bit stream and redundancy is removed; thus, data compression is realized. Subsequently, channel encoding is employed to introduce redundancy into the generated bit stream to resist channel noise. Then, the bitstream is modulated into symbols to further improve the transmission efficiency. After transmission through the communication channel, the symbols are demodulated and decoded for measurement recovery. Finally, computational reconstruction algorithms are applied over the received measurement for information recovery. However, the straightforward combination (SC) scheme of digital communication and CI is not optimal because of the following technical gap: First, the communication systems ignore the formation of the optical measurements. Different from natural images, the optical measurements contain both the optical information and optical transfer function. Hence, irreversible errors are introduced by applying conventional image source coding algorithms. Second, the CI systems lack resistance to communication channel disturbance. When passing through the communication systems, the measurement is degraded by channel noise and other impairments, making it hard for the corresponding computational reconstruction algorithms to perform the subsequent recovery. In addition to the technical gap, another urgent issue to be solved is the slow response of the SC scheme because of trivial data processing procedures.

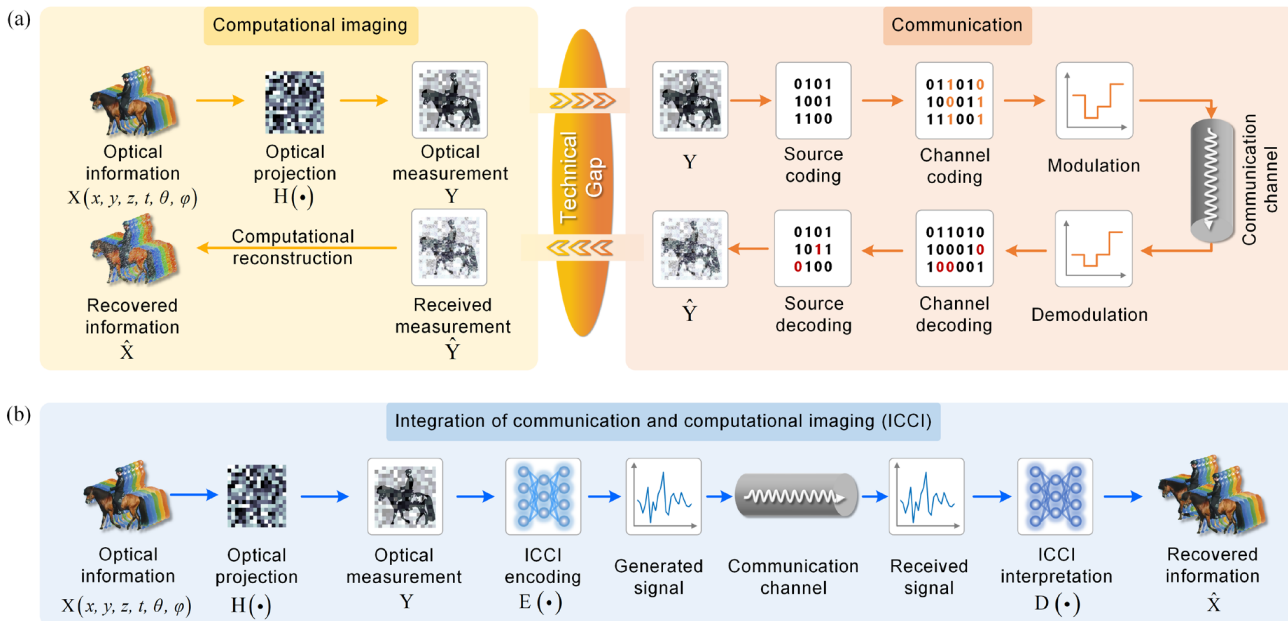

**Fig. 1.** Remote perception pipeline achieved by (a) straightforward combination (SC) with communication and computational imaging and (b) integration of communication and computational imaging (ICCI) framework. $X(x,y,z,\lambda,t,\theta,\varphi)$ is the light field function. $(x, y, z)$ denotes the spatial position. $\lambda$ is the wavelength and $t$ is the time. $(\theta, \varphi)$ denotes the direction of light.

To overcome the cross-domain impairments and bridge the technical gap, we propose the ICCI paradigm for highly efficient remote perception. Fig. 1(b) depicts the data flow of the ICCI scheme. Intrinsically, traditional separable coding methods attempt to realize data compression and high-fidelity transmission. To simplify this process, we utilize an ICCI encoder to directly map the optical measurement to the compressed, discrete-time analog symbols. At the receiver, an ICCI interpreter is designed to directly recover the remote visual information from the one-dimensional received symbols. Essentially, optical encoding and digital encoding perform linear projections on the visual information twice. Hence, we utilize one ICCI interpretation process to realize all inverse projections to recover the visual information. Both the ICCI encoder and interpreter are constructed using deep convolutional neural networks. To realize transmission robustness against channel noise, the ICCI encoder–interpreter network is trained under a given channel model. Benefiting from the end-to-end learning advantages, the ICCI network achieves full-link optimization by considering the final vision tasks and the disturbance induced by the communication channel. Suppose that the original optical information is $X$, then the optical measurement captured by the front optics can be expressed as $Y = H(X)$, where $H$ is the forward model of the CI systems. Let $E_\alpha(\cdot)$ and $D_\beta(\cdot)$ denote the ICCI encoder and the ICCI interpreter, respectively, where $\alpha$ and $\beta$ are the learnable parameter sets. Considering the communication channel noise $N(\cdot)$, the interpreted visual information can be expressed as $\widehat{X} = D_\beta \left( N \left( E_\alpha(Y) \right) \right)$. By jointly considering the acquisition and transmission phases, the ICCI framework aims to minimize the information loss function:

$$\underset{\alpha,\beta}{\mathrm{argmin}}\, Loss_{task}\left( D_\beta \left( N \left( E_\alpha \left( H(X) \right) \right) \right), X \right), \quad (1)$$

supervised by the loss function, the ICCI network optimizes $\alpha$ and $\beta$ through the gradient descent method [30]. Note that the loss function is task-oriented, which means different constraints can be introduced according to different vision tasks. For instance, in reconstruction tasks, such as temporal imaging [31-32] and spectral imaging [33-34], pixel-wise mean square error (MSE) is commonly adopted as the loss function. By designing different loss functions, the ICCI framework is expected to execute various downstream remote CV tasks, such as applying intersection over union (IoU) loss for object detection [35] and applying cross-entropy (CE) loss for object classification [36].

III. NUMERICAL ANALYSIS OF REMOTE HYPERSPECTRAL INFORMATION PERCEPTION

Hyperspectral images contain the spectral features of the object and are widely applied in military security [37], environmental monitoring [38], and so forth. In this section, we conduct numerical analysis with the ICCI framework on remote hyperspectral information perception. The designed ICCI system could capture 27-bands hyperspectral images with a spatial resolution of $256 \times 256$ from 450 nm to 700 nm.

*A. System design*

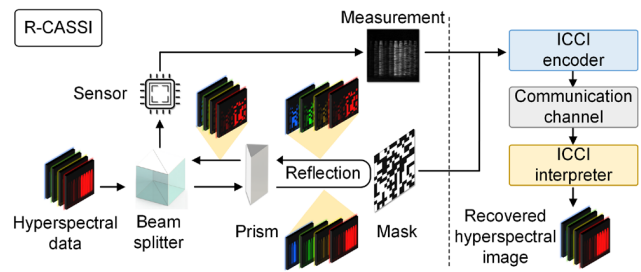

**Fig. 2.** Workflow of remote hyperspectral information perception with the ICCI framework.

Fig. 2 shows the system design of remote hyperspectral information perception achieved by using the ICCI framework. In the optical encoding phase, a typical CI method, that is, the CASSI method, is employed to capture the optical measurement. Particularly, we employ the reflective CASSI (R-CASSI) method [28], here by considering the compactness of the front-end optics. The three-dimensional hyperspectral image can be described as $X(x, y, \lambda)$. In the R-CASSI system, after dispersion by a prism and modulation by an optical encoding element (i.e., mask), the hyperspectral data are detected as follows:

$$Y(x, y) = \int_{\lambda_{min}}^{\lambda_{max}} X(x, y, \lambda) \cdot M(x, y, \lambda) \cdot d\lambda + G(x, y), \quad (2)$$

$M(x, y, \lambda)$ denotes the pattern of the mask and $G(x, y)$ is the noise term. $[\lambda_{min}, \lambda_{max}]$ represents the working band of the R-CASSI system. According to compressive sensing theory [39], (2) can be rewritten in a matrix form as:

$$y = Hx + g, \quad (3)$$

where $y \in \mathbb{R}^{(mn) \times 1}$ is the measurement and $m, n$ are the pixelated spatial coordinates. $x \in \mathbb{R}^{(kmn) \times 1}$ is the spectral data, where $k$ denotes the number of discrete spectral bands. $H \in \mathbb{R}^{(mn) \times (kmn)}$ is the sensing matrix that can be obtained from the pattern of the mask. $g \in \mathbb{R}^{(mn) \times 1}$ is the discretized noise matrix. By applying computational reconstructions algorithms, the hyperspectral information $x$ can be recovered from the measurement $y$.

Then, an ICCI encoder turns the measurement into a symbol sequence for transmission. To simulate the communication channel environment, we consider two widely used channel models: the AWGN channel model and slow fading channel model. The communication channel transfer function is modeled as follows:

$$\hat{s} = \begin{cases} s+n, & (AWGN\ channel) \\ h\cdot s+n, & (slow\ fading\ channel) \end{cases}, \quad (4)$$

where $s$ is the transmitted data and $\hat{s}$ is the received data. $n$ is the noise vector sampled from Gaussian distribution. $h$ is a normal random variable denoting the channel gain in a slow fading channel. At the receiver, the received symbol sequence is recovered to the hyperspectral information by an ICCI interpreter.

*B. Implementation of the ICCI encoder and ICCI interpreter*

Fig. 3(a) shows the structures of the ICCI encoder and ICCI interpreter. All modules deployed in the ICCI network use convolution as basic computation operations. The feature projection layer utilizes four scales of convolution kernels to preliminarily extract useful information from the optical measurement. Fig. 3(b) shows the structure of the triple residual block (TRB) layer. The TRB layer utilizes three residual connections to realize the fusion of deep and shallow information. Fig. 3(c) shows the details of the spatial-spectral attention (SSA) layer, which obtains the spatial and spectral correlations of the hyperspectral information. Fig. 3(d) shows the details of the importance-aware layer (IAL), which protects significant information against communication channel noise. The upsampling layer expands the spatial resolution of the input data by two times.

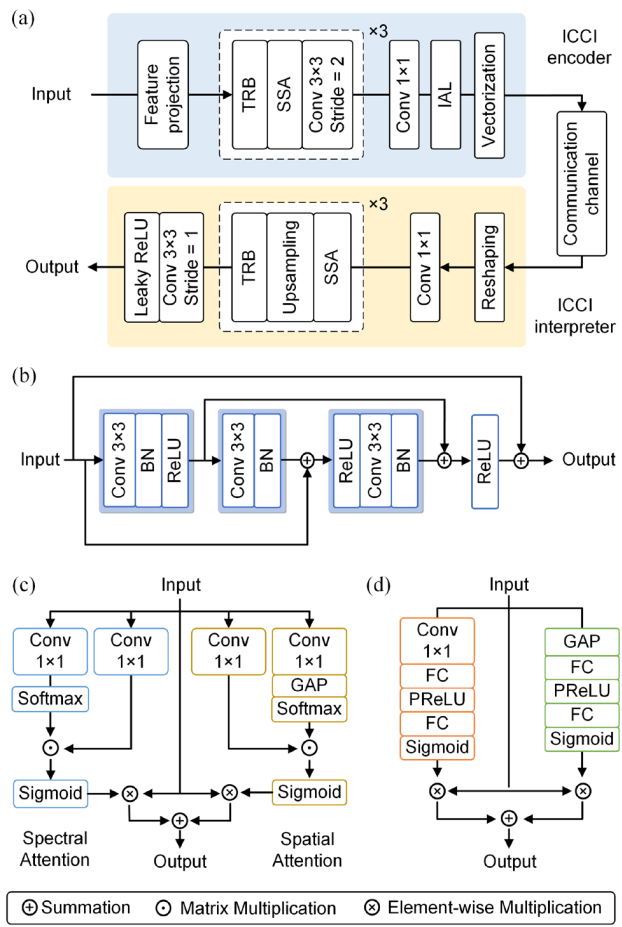

**Fig. 3.** (a) Structures of the ICCI encoder and ICCI interpreter. (b) Triple residual block. BN: batch normalization. ReLU: rectified linear unit. (c) Spatial-spectral attention layer. GAP: global average pooling. (d) Importance-aware layer. FC: fully connected layer.

To train the ICCI network, we use MSE as the loss function by considering the final hyperspectral reconstruction task. The MSE loss is expressed as:

$$MSE = \frac{1}{N}\left\|\hat{X}-X\right\|_2^2, \quad (5)$$

where $\|\cdot\|_2$ is the $L_2$ norm. $\hat{X}$ is the recovered data and $X$ is the ground truth. $N$ is the total number of pixels in the hyperspectral image. During the training process, channel noise is added to the output of the ICCI encoder. The ICCI network is trained on the CAVE hyperspectral dataset [40] and validated on the Kaist hyperspectral dataset [41].

*C. System Performance Evaluation*

For performance comparison, we also conduct SC schemes by combining digital communication systems and the R-CASSI system. JPEG2000 [13] is employed for source coding. In this case, 2/3, 3/4 rate low-density parity-check code (LDPC) [14] is adopted for channel coding. As for the modulation formats, we apply PAM4 and PAM8 for intensity modulation direct detection (IM-DD) transmission systems, while BPSK, QPSK, and 16QAM are applied for coherent transmission systems. The transmission rates of these systems are kept consistent. In the SC schemes, Unet-3D algorithm [28] is used for computational reconstruction.

Additionally, we define the joint data compression ratio (DCR) of the remote hyperspectral perception systems as:

$$DCR = \frac{L}{H\times W\times C} \quad (spv), \quad (6)$$

where $H\times W\times C$ is the shape of the hyperspectral images. $L$ is the number of symbols generated with the ICCI encoder. Considering that hyperspectral images are three-dimensional data composed of voxels, we define the unit of DCR as a symbol per voxel (spv). Here, the DCRs of these numerical analysis schemes are set to 0.005 spv.

To quantitatively evaluate the system's performance, peak signal-to-noise ratio (PSNR) and structural similarity index matrix (SSIM) [42] are calculated between the recovered and ground truth hyperspectral images. A higher PSNR or SSIM value indicates a better result. Furthermore, we calculate the spectral errors (SE) between the reconstructed spectrum and the ground truth spectrum. For a hyperspectral data with a shape of $H\times W\times C$, the spectrum of a spatial point can be described as a spectral vector with a shape of $1\times 1\times C$. The SE is expressed as:

$$SE = \sum_{i=1}^{n}(v_i - \hat{v}_i)^2, \quad (7)$$

where $v \in \mathbb{R}^C$ is the ground truth spectrum and $\hat{v} \in \mathbb{R}^C$ is the recovered spectrum. A lower SE value indicates a better spectral recovery.

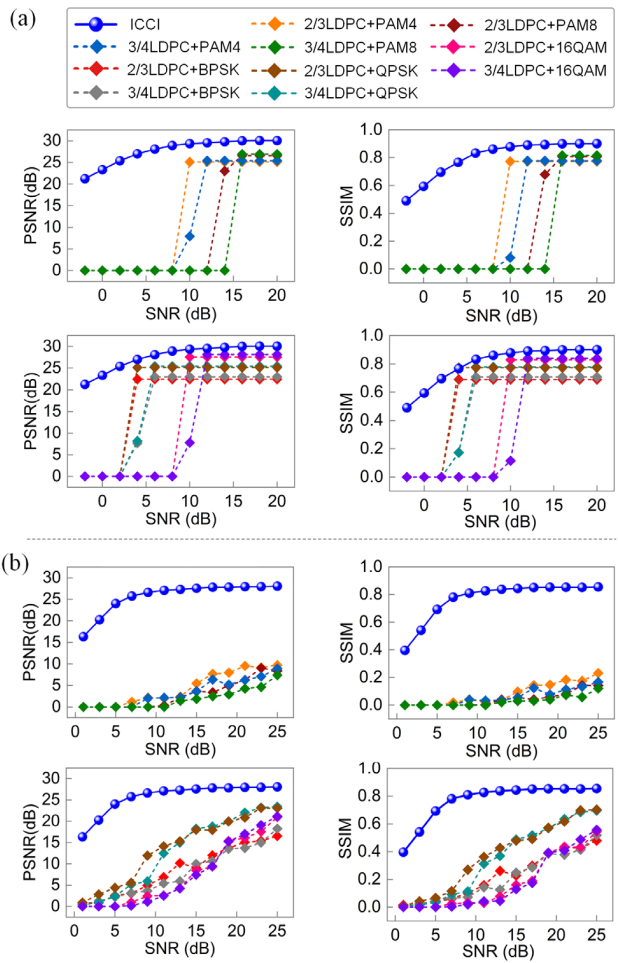

**Fig. 4.** Performance with respect to the SNR in (a) AWGN channel transmission and (b) slow fading channel transmission.

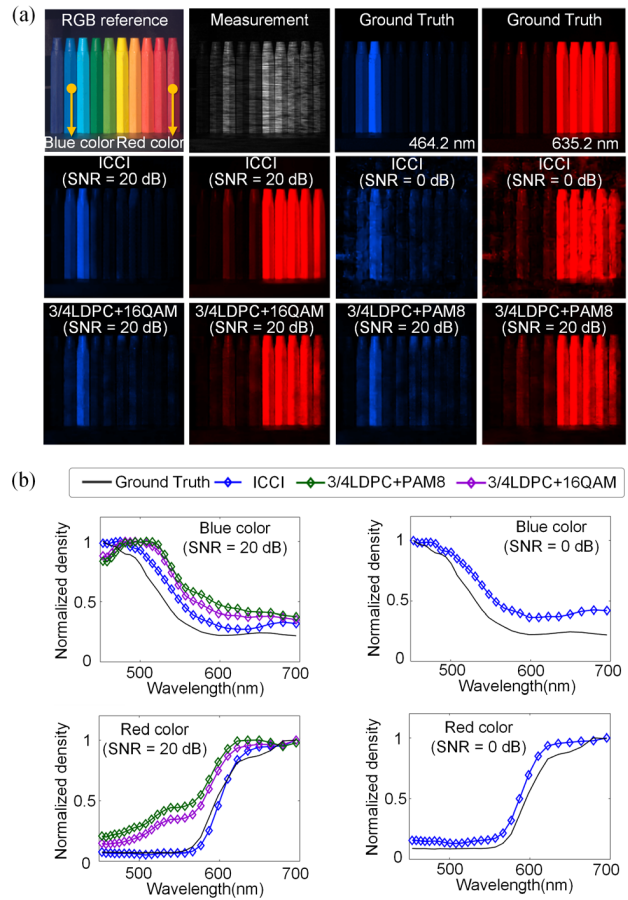

**Fig. 5.** Spatial and spectral evaluation in the AWGN channel transmission. (a) Visual comparison of the recovered two-band spectral images. (b) Comparison of spectral recovery precision.

Fig. 4 shows the results evaluated over 10 Kaist hyperspectral images with respect to the signal-to-noise ratio (SNR) in AWGN channel transmission (Fig. 4(a)) and slow fading channel transmission (Fig. 4(b)). In AWGN channel transmission, the ICCI scheme is more robust to additive noise. The SC schemes experience drastic performance degradation, where the received data are severely impaired by the channel noise so that the source decoding algorithms fail to recover the optical measurements. In slow fading channel transmission, the ICCI scheme shows stronger robustness against channel fluctuations compared with the SC schemes.

Fig. 5 shows the spatial and spectral evaluation in AWGN channel transmission. The results of two SC schemes (i.e., 3/4LDPC+16QAM, 3/4LDPC+PAM8) are selected for comparison. Fig.5(a) plots the recovered two-band spectral images (464.2 nm, 635.2 nm). When the SNR is 20 dB, the ICCI scheme achieves better spatial recovery compared with the SC schemes. When the SNR decreases to 0 dB, the SC schemes fail to recover the hyperspectral information, while the ICCI scheme still achieves an acceptable performance. In Fig. 5(b), we plot the spectral curves of the blue and red colors (marked on the RGB reference image in Fig. 5(a)). We calculate the SE values between the recovered spectral curves and ground truth spectral curves for the three schemes. For the blue color, the ICCI scheme achieves a 0.303 SE value when the SNR is 20 dB. However, the 3/4LDPC+16QAM and 3/4LDPC+PAM8 only achieve 1.406 SE value and 1.529 SE value, respectively. When the SNR decreases to 0 dB, the ICCI scheme still achieves a 0.527 SE value, whereas the SC schemes fail to recover the spectral information. The analysis of the red color also provides similar results. The results further indicates that the ICCI scheme enables a more accurate and stable spectral recovery compared with the SC schemes.

Fig. 6 shows the spatial and spectral evaluation in slow fading channel transmission. We select the 2/3LDPC+QPSK and 2/3LDPC+PAM4 methods for comparison. Fig. 6(a) shows the recovered spectral images when the SNRs are 25 dB and 5 dB. We observe that the spatial details of the hyperspectral object are severely impaired after applying the SC schemes. However, the ICCI scheme achieves high-fidelity recovery. Fig. 6(b) plots the spectral curves of the three methods. For the blue color, the ICCI scheme achieves a 0.499 SE value when the SNR is 25 dB. However, the 2/3LDPC+QPSK and 2/3LDPC+PAM4 only achieve SE values of 3.378 and 4.272, respectively. When the SNR decreases to 5 dB, the ICCI scheme achieves a 1.055 SE value, whereas the 2/3LDPC+QPSK and 2/3LDPC+PAM4 achieve SE values of 4.681 and 7.847, respectively. The analysis of the red color also provides similar results. The results further indicate that the ICCI scheme is more robust to channel fluctuations compared with the SC schemes.

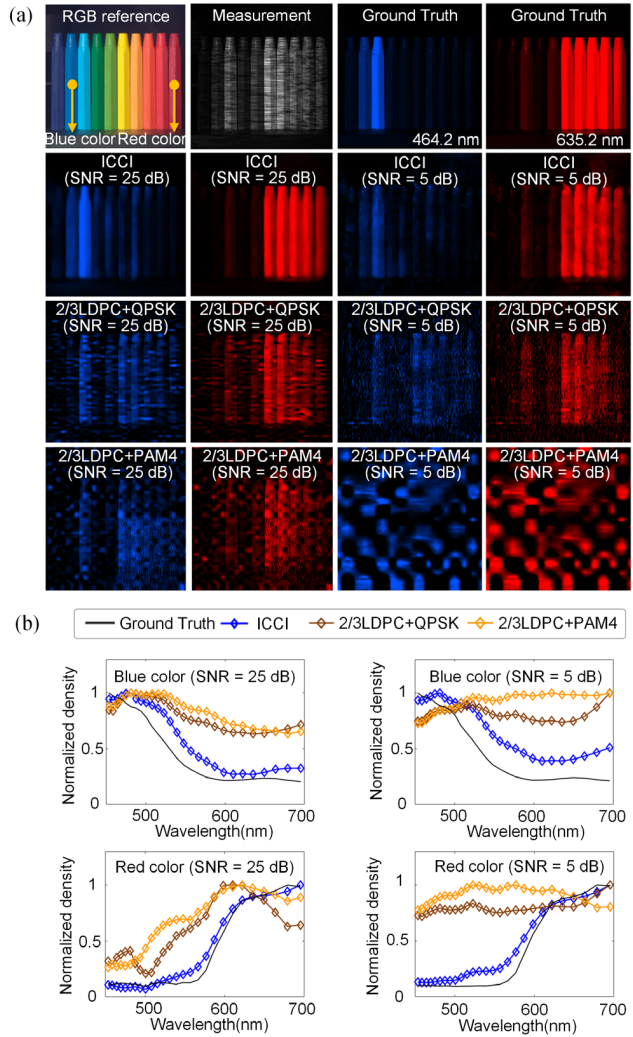

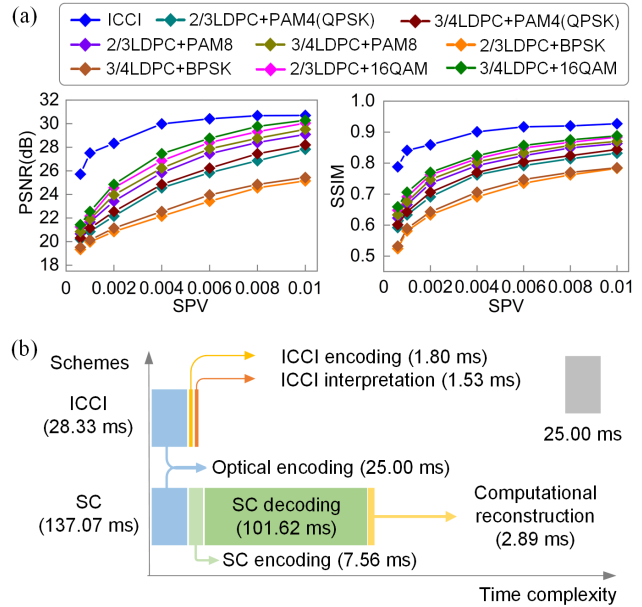

**Fig. 6.** Spatial and spectral evaluation in the slow fading channel transmission. (a) Visual comparison of the recovered two-band spectral images. (b) Comparison of spectral recovery precision.

**Fig. 7.** (a) Performance of different schemes with respect to the DCR. (b) Time complexity of the ICCI scheme and the SC scheme.

Fig. 7(a) shows the performance of the ICCI scheme and SC schemes under different DCRs (0.0006, 0.001, 0.002, 0.004, 0.006, 0.008, and 0.01). The transmission rates of these systems are kept consistent. The results are obtained in AWGN channel transmission with 20 dB SNR. Under this SNR, the SC systems could achieve the optimal performance, which refers to that the bitstream can be transmitted without errors. Therefore, the optimal performance difference of the SC schemes is determined by the compression ratios of lossy source coding. Because of the same source compression ratios, the PAM4 and QPSK systems can achieve the same PSNR and SSIM performance, which can be represented by one curve. The results illustrate that the ICCI scheme can realize a higher data transmission rate with the same performance compared with the SC schemes.

Fig. 7(b) shows the time complexity comparison between the ICCI scheme and SC scheme. We monitor the time cost for each step in both schemes. Particularly, we adopt the JP2K+2/3LDPC+PAM4 method for comparison by considering its lowest time complexity among all the SC schemes. The analysis is conducted on a machine equipped with an Intel Xeon Gold 5218 CPU and two NVIDIA RTX 3090 GPUs. For a $256 \times 256 \times 27$ size hyperspectral image, the total time cost of the ICCI scheme and SC scheme are 28.33 ms and 137.07 ms, respectively. The lower time complexity of the ICCI scheme results from full-link optimization, which enables the ICCI system to achieve better performance with a more concise structure.

## IV. NUMERICAL ANALYSIS OF REMOTE HIGH-SPEED VIDEO INFORMATION PERCEPTION

High-speed video information is a significant tool in object detection [43] and collision analysis [44] because of its high temporal resolution. In this section, we conduct numerical analysis with the ICCI framework on remote high-speed video information perception. The designed ICCI system could capture 500 fps high-speed videos with a spatial resolution of $256 \times 256$.

### A. System design

Fig. 8 shows the system design of remote high-speed video information perception achieved by using the ICCI framework. We choose the coded aperture compressive temporal imaging (CACTI) system to achieve the optical projection because of its ability to acquire high-speed video with a low frame rate camera [45]. In the CACTI system, the high-speed video is modulated by a digital mirror device (DMD) and then compressed into a single measurement. By adopting computational reconstruction, the high-speed video information can be recovered from the single shot. Let $X(x, y, t)$ denote the video data. $t$ is the time. The mathematical model of CACTI is expressed as follows:

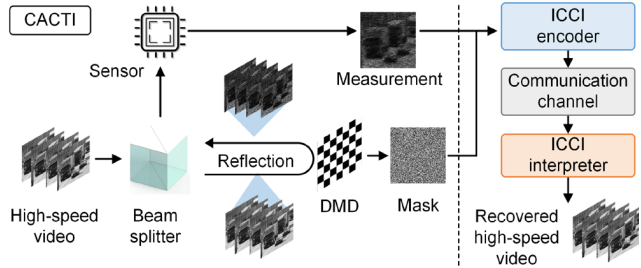

**Fig. 8.** Workflow of remote high-speed video information perception with the ICCI framework.

$$Y(x, y) = \int_{t_0}^{t} X(x, y, t) \cdot M(x, y, t) \cdot dt + G(x, y), \quad (8)$$

where $Y(x, y)$ is the captured measurement. $M(x, y, t)$ is the modulation information (i.e., mask) performed by the DMD at $t$ moment. $[t_0, t]$ is the integration time of the sensor. $G(x, y)$ denotes the noise. According to compressive sensing theory, (8) can be rewritten in the same form as (3).

The right part of Fig. 8 shows the transmission and interpretation process of the high-speed video information. Here, we use the same channel models as in the section III.

### B. Implementation of the ICCI encoder and ICCI interpreter

Fig. 9 shows the implementation details of the ICCI network for high-speed video perception. The initialization layer uses the optical measurement and mask information as input to ensure effective information extraction. In the ICCI network, we apply the residual dense net (ResDnet) block [46] as basic module. This block enables accurate extraction of abundant video features. The three-dimensional convolution (Conv3D) is adopted to obtain the spatial and temporal correlations of the video's information. The ICCI network is trained on the DAVIS 2017 dataset [47] and validated on a benchmark video dataset [48]. MSE is applied as the loss function for the network training.

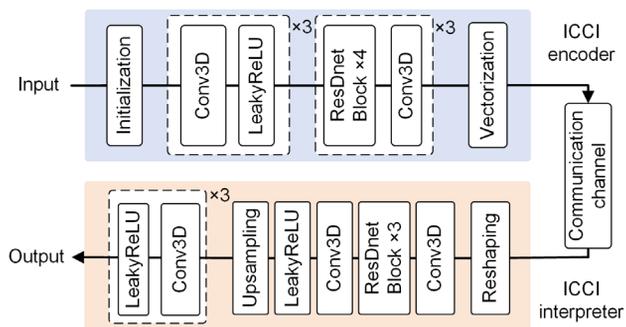

**Fig. 9.** Structure of the ICCI encoder–interpreter network. Conv3D: three-dimensional convolution. ResDnet: residual dense net. LeakyReLU: leaky rectified linear unit.

### C. System Performance Evaluation

For performance comparison, we adopt the same SC transmission schemes as in the section III. The EfficientSCI algorithm [46] is applied for computational reconstruction.

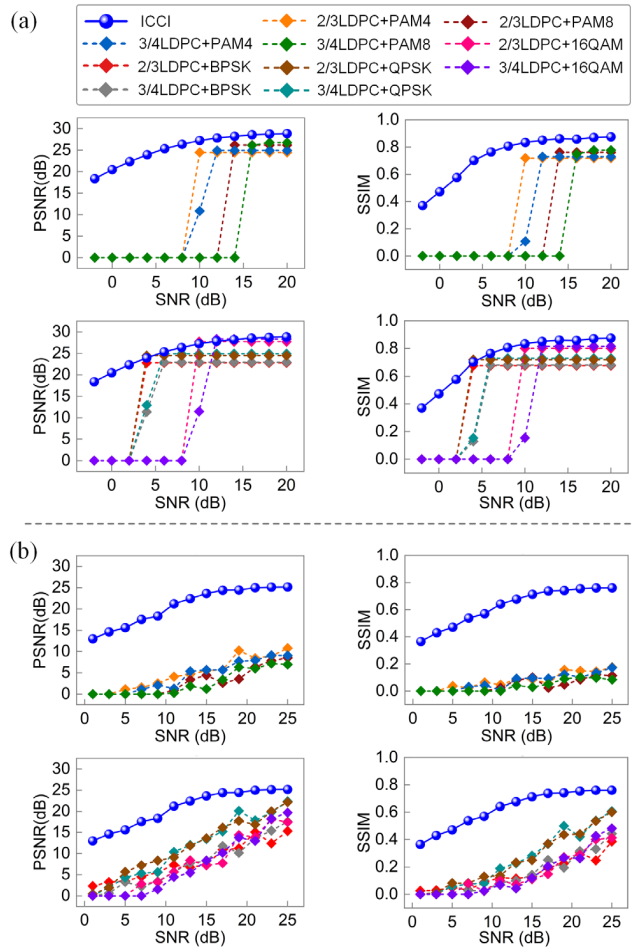

**Fig. 10.** Performance evaluation with respect to the SNR in (a) AWGN channel transmission and (b) slow fading channel transmission.

The DCRs of all schemes are set to 0.031 spv. The transmission rates of these systems are kept consistent. PSNR and SSIM are calculated for quantitative evaluation.

Fig. 10 shows the results evaluated over six high-speed videos in AWGN channel transmission (Fig. 10(a)) and slow fading channel transmission (Fig. 10(b)). We can observe the same phenomenon as in the hyperspectral perception analysis. The ICCI scheme shows great robustness against additive noise and channel fluctuations compared with the SC schemes.

Fig. 11(a) shows the visualization of the recovered videos in AWGN channel transmission. When the SNR is 20 dB, the ICCI scheme achieves better spatial recovery and avoids the artifacts that appear in the SC schemes. When the SNR decreases to 0 dB, the ICCI scheme ensures acceptable performance, whereas the SC schemes fail to recover the video information. Fig. 11(b) shows the visualization in slow fading channel transmission. When the SNR is 25 dB, the ICCI scheme recovers spatial information more accurately. When the SNR decreases to 5 dB, the main features of the video information are recognizable by applying the ICCI scheme. However, by applying SC schemes, the video information is severely impaired.

The analysis of the remote high-speed video information perception further verifies that the ICCI framework could handle with different types of high-dimensional optical information.

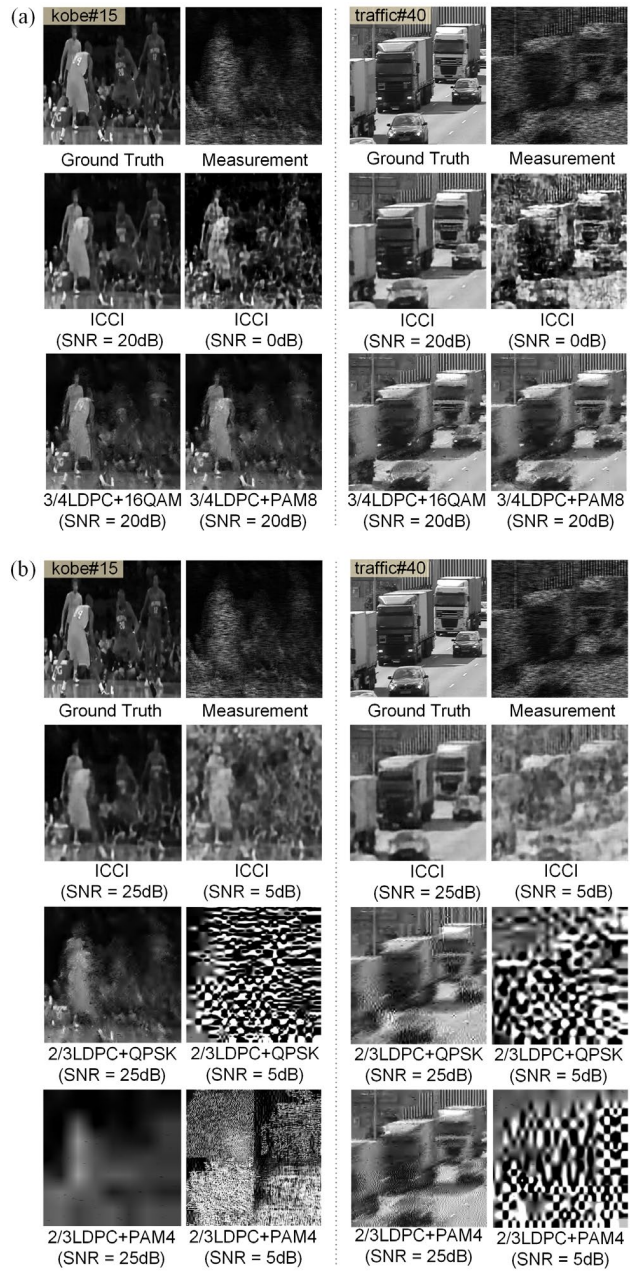

**Fig. 11.** Visual comparison of two recovered high-speed videos in (a) AWGN channel transmission and (b) slow fading channel transmission. #N denotes the Nth frame of the videos.

## V. EXPERIMENTAL DEMONSTRATION OF REMOTE HYPERSPECTRAL INFORMATION PERCEPTION

To illustrate the feasibility of the ICCI framework, we physically implement the experimental setup with the IM-DD optical fiber transmission system and R-CASSI system, as shown in Fig. 12(a). With the ICCI framework, the remote hyperspectral information is sequentially encoded by the R-CASSI system and ICCI encoder into discrete symbols. Then, the symbols are sent into the IM-DD system for data transmission. At the receiver, we design the channel distortion aware network (CDAN) to remove the channel distortion. After being equalized by the CDAN, the received symbols are interpreted into hyperspectral information. For performance comparison, we adopt two traditional SC schemes, i.e., JPEG2000+2/3LDPC+PAM4+Unet-3D and JPEG2000+2/3LDPC+PAM8+Unet-3D in the experiments. The 61-tap feed-forward equalization (FFE) is applied for channel equalization in the SC schemes. The DCRs of these schemes are set to 0.005 spv. The symbol rate of the transmission systems is 30 Gbaud. The transmission rates of the SC schemes and the ICCI scheme are kept similar, i.e., $2.827 \times 10^5$ images/s and $2.828 \times 10^5$ images/s, respectively. Considering that a hyperspectral image contains $1.699 \times 10^8$ bits, the transmission rate of this ICCI system could reach 48 Tbit per second.

### A. Implementation details

The optical setup of R-CASSI system is shown in Fig. 12(b). An objective lens (M1614-MP2, Computar) relays the hyperspectral scene onto an intermediate plane, which is located at the focal point of relay lens 1 (ACA254-060-A, Thorlabs). After being dispersed by the prism (PS811-A, Thorlabs), the dispersed spectral data are subsequently relayed by lens 2 (ACA254-060-A, Thorlabs) onto the mask plane for modulation. The mask is a binary random pattern, where "1" patterns reflect the incident light and "0" patterns transmit the incident light. Encoded and reflected by the mask, the spectral data are inversely dispersed by the prism and detected by a monochromatic camera (acA2000-165um, Basler) after passing through the beam splitter (BS013, Thorlabs). Two filters (FELH0450 and FESH0700, Thorlabs) are applied to limit the working band of the R-CASSI system to 450–700 nm. Determined by the parameters of the hardware, 27-band spectral data ranging from 450 nm to 700 nm are captured by the R-CASSI system.

In the IM-DD transmission system, the symbols generated by the ICCI encoder are first processed by resampling, square-root-raised cosine and digital pre-emphasis. Then the symbols are sent into an arbitrary waveform generator (AWG, M8195A, Keysight). The highest sampling rate and analog bandwidth are 65 GSa/s and 25 GHz, respectively. The electrical signals generated by the AWG are then amplified by a linear broadband amplifier (SHF S807C) and modulated into optical signals by a Mach–Zehnder modulator (MZM, AX-0MVS-40-PFA-PFA-LV, EOSPACE). In optical back-to-back transmission, a variable optical attenuator (VOA) is applied to control the ROP. In fiber transmission, the input fiber optical power and ROP are both 0 dBm, with the control of the VOA and erbium-doped fiber amplifier (EDFA). At the receiver, the optical signals are converted to electrical signals by a photodetector (PD, XPRV2325A, FINISAR) with a 30 GHz bandwidth. The obtained electrical signals are sampled at 100 GSa/s by a digital sampling oscilloscope (DSO, DSA72504D, Tektronix) with an analog bandwidth of 25 GHz. Moreover, resampling is applied to upsample to two samples per symbol.

Fig. 12(c) shows the details of the CDAN. The input symbol sequence contains 121 symbols, with $s_n$ located at the middle. The first layer and second layer contain 60 neurons and 121 neurons, respectively. After removing the distortion of symbol $s_n$, the CDAN outputs the equalized symbol $z_n$. In the experiments, the equalization process is divided into two steps. First, we utilize 5% of the received symbols to train the CDAN. The mean square error (MSE) is used as the loss function. After the training step, the parameters of the CDAN are fixed. By inputting the total received symbols, the channel distortion is removed. Then, the equalized symbols are sent to the ICCI interpreter for hyperspectral information interpretation.

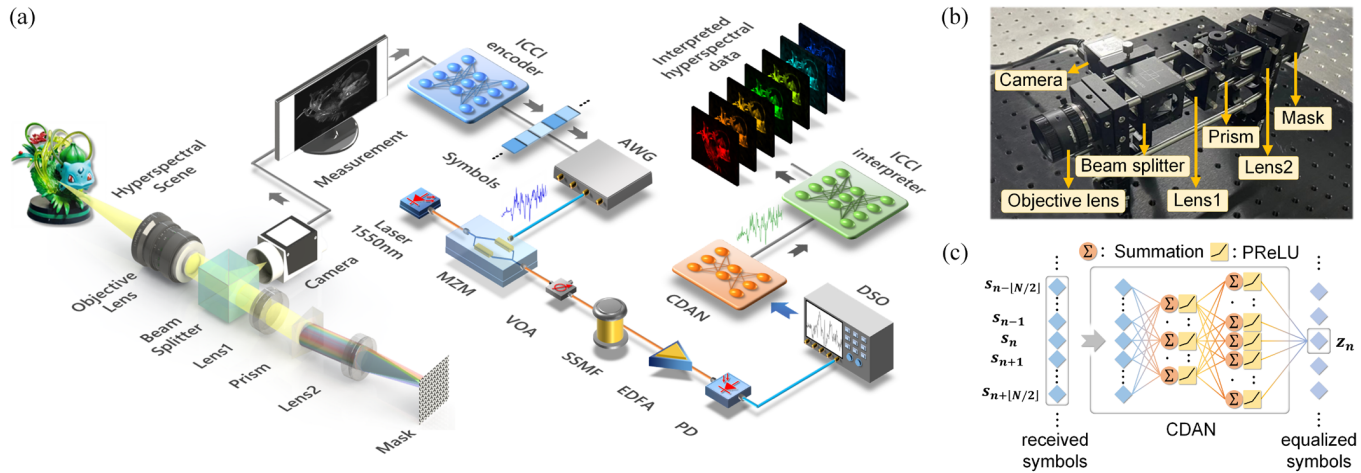

**Fig. 12.** (a) Experimental setup with the IM-DD optical transmission system and R-CASSI system. (b) Optical setup of R-CASSI system. (c) Structure of the CDAN. PReLU: parametric rectified linear unit.

Fig.13 shows the recoverd 27-band spectral images when the ROP is 0 dBm and the transmission distance is 30 km.

### B. Optical back-to-back transmission experiment

In the optical B2B transmission experiment, we validate the system performance from -14 dBm to 0 dBm with a 1 dBm step. Fig. 14(a) shows the RGB reference and optical measurement of the hyperspectral scene. Three regions are selected for performance evaluation. We calculate the SE values between the ground truth spectral curves and recovered spectral curves of the blue (in the first region), green (in the second region), and red spectra (in the third region). The ground truth spectral curves are obtained with a commercial spectrometer (FLAME-T-UV-VIS-ES, ocean insight). Fig. 14(b) shows the results according to the different ROPs. Consistent with the numerical analysis, the SC schemes experience cliff-like degradation when the ROP is -4 dBm for PAM8 and -8 dBm for PAM4. However, the ICCI scheme achieves a stable spectral recovery, even in a low ROP regime. Fig. 14(d) plots the reconstructed characteristic spectral images (blue: 481.5 nm, green: 530.8 nm, red: 635.2 nm) of three marked regions. When the ROP is 0 dBm, the ICCI scheme achieves smoother recovery, whereas some artifacts appear by adopting the SC schemes. When the ROP is -10 dBm, the SC schemes fail to recover the information. By applying the ICCI scheme, the hyperspectral information is successfully recovered with acceptable performance degradation. Therefore, the ICCI system can realize remote optical information perception at higher noise levels.

### C. Optical fiber transmission experiment

In the fiber transmission experiment, we validate the system performance under 11 transmission distances from 0 to 80 km. Both the PAM4 and PAM8 systems are deployed without special dispersion compensation. Fig. 14(c) shows the spectral recovery precision under different transmission distances. The PAM4 and PAM8 systems only reliably transmit the information within 10 km, whereas the ICCI scheme provides an acceptable performance over 80 km. Fig. 14(e) displays the spectral images of the three marked regions. When the transmission distance is 5 km, the ICCI scheme avoids the artifacts that appear in the SC schemes. When the transmission distance reaches 80 km, the SC schemes fail to recover the hyperspectral information. The ICCI scheme still achieves acceptable visual performance and reliable spectral recognition. The stable spectral recovery illustrates that the ICCI scheme can achieve optical information perception over long distances.

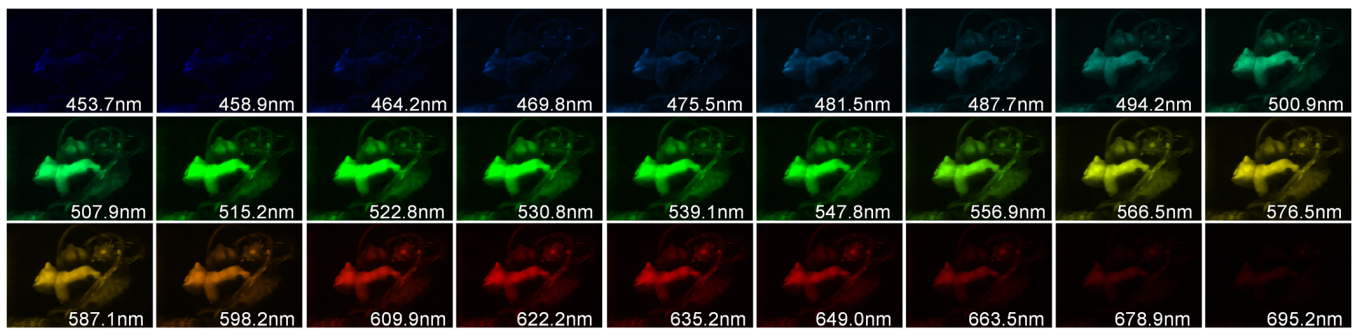

**Fig. 13.** Visualization of 27-band recovered hyperspectral image. The ROP is set to 0 dBm and the transmission distance is set to 30 km.

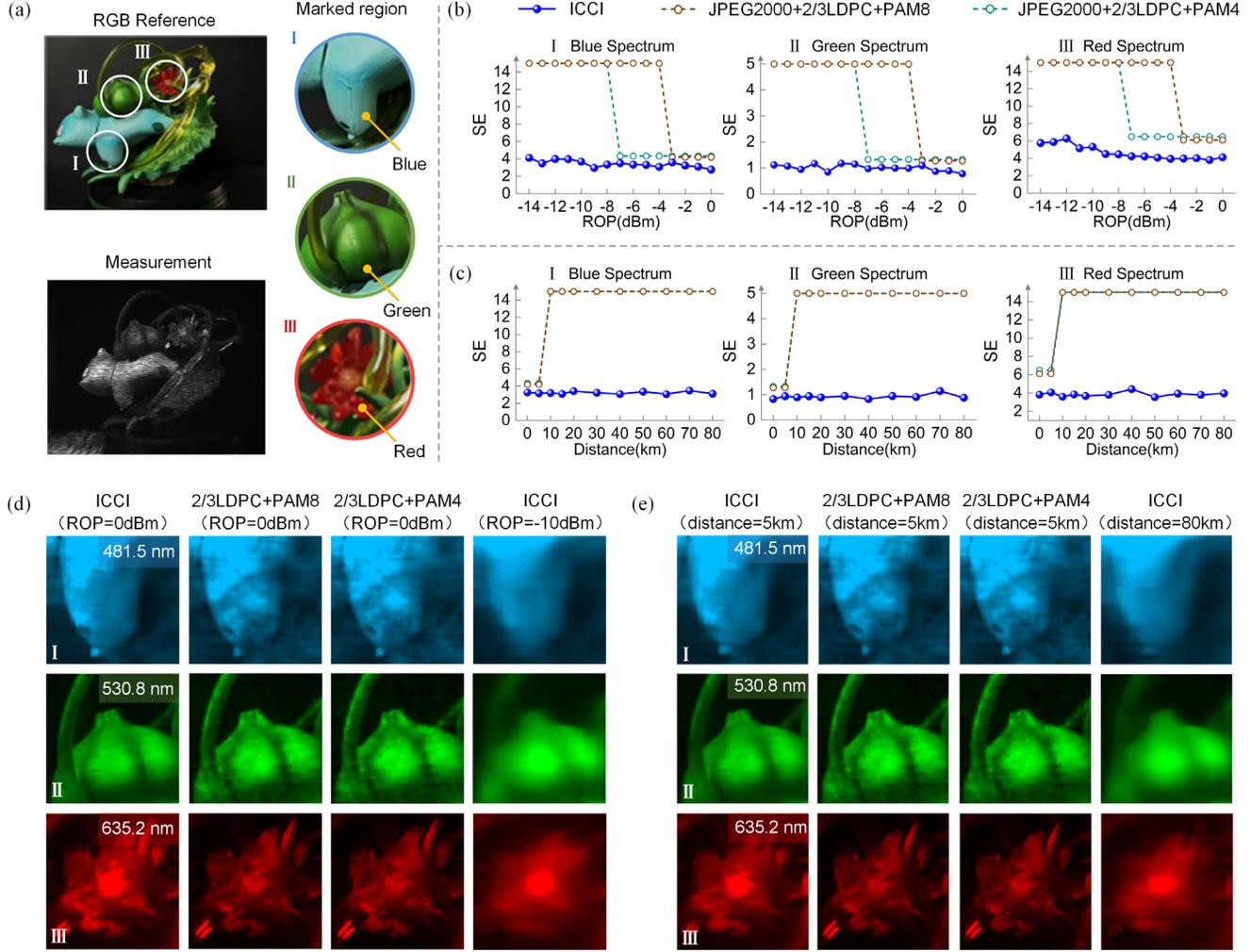

**Fig. 14.** (a) Hyperspectral scene and optical measurement. Spectral recovery precision under different (b) ROPs and (c) distances. Recovered three-band spectral images of the marked regions when (d) ROPs are 0 dBm and -10 dBm, and (e) distances are 5 km and 80 km.

*D. Fast remote four-dimensional hyperspectral video perception*

Furthermore, we realize remote hyperspectral video perception with the ICCI experimental setup. In this experiment, the ROP is set to 0 dBm, and the transmission distance is set to 80 km. The R-CASSI system captures 60 optical measurements in 2 seconds with a frame rate of 30 frames per second. To increase the data transmission rate, we set the DCR to 0.002 spv. In this case, the ICCI system could transmit $7.06 \times 10^5$ hyperspectral images per second. The transmission rate reaches 120 Tbit per second.

Fig. 15 shows the optical measurements and the recovered three-band spectral images of four video frames. The spatial details of the hyperspectral object can be easily recognized. Fig. 16(a) calculates the SE values of the green spectrum in 60 video frames. We can see that the SE value fluctuates at a lower level, indicating the high stability of the ICCI system with time variation. Fig. 16(b) plots the recovered and ground truth spectral curves of the green spectrum in four frames. Near the 530 nm wavelength, which is the characteristic band representing the green color, the recovered spectrum has a very close peak to the ground truth spectrum peak that is measured by the spectrometer, verifying the reliable spectrum identification in video-level remote hyperspectral information perception. The average encoding time and interpretation time of each frame are 5.13 ms and 2.28 ms, respectively. The average equalization time of each frame is 0.63 ms. For the acquisition part, the existing commercial camera can capture images beyond a 30fps speed. Therefore, the ICCI scheme could realize remote optical information perception with low latency. This reveals the potential to apply ICCI system to execute massive remote real-time CV tasks.

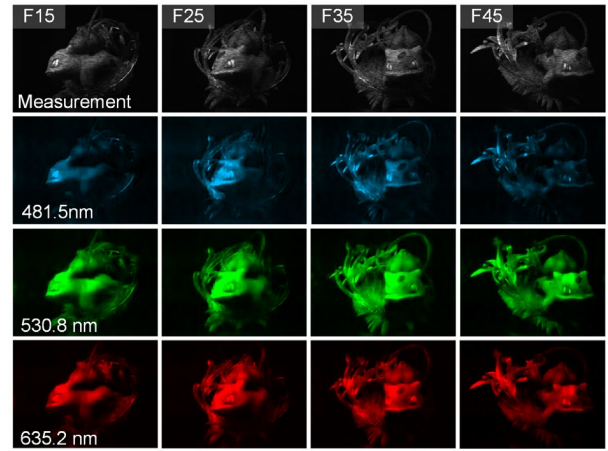

**Fig. 15.** Optical measurements and recovered three-band spectral images of four hyperspectral video frames.

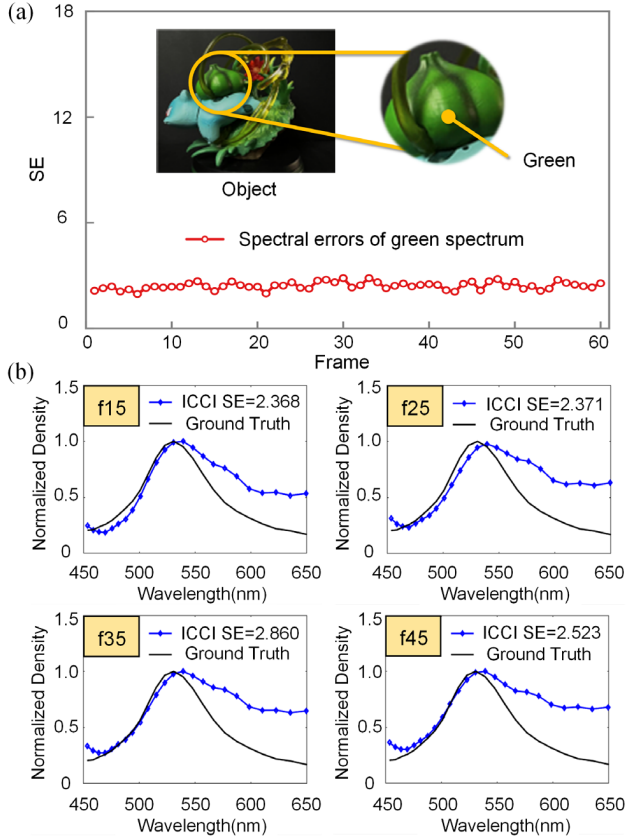

**Fig. 16.** (a) SE values of the green spectrum in 60 video frames. (b) Spectral curves of the green spectrum in four frames. fN denotes the Nth frame.

## VI. EXPERIMENTAL DEMONSTRATION OF REMOTE HIGH-SPEED INFORMATION PERCEPTION

In this section, we conduct analysis on the real high-speed video information. The experimental system could capture 500 fps high-speed videos with a spatial resolution of $330\times330$. The optical setup of CACTI system is shown in Fig.17. An objective lens (LM50HC, Kowa Lens) relays the high-speed scene onto the DMD plane (V-7001, VIALUX). Then, the DMD modulates the video information with random "0" and "1" patterns. Reflected by the DMD, the encoded video information sequentially passes through a beam splitter (CCM1-BS013, Thorlabs), relay lens 1 (TTL100-A, Thorlabs), and relay lens 2 (C23-3520-2M, Basler). Finally, a monochromatic camera (acA1300-220um, Basler) is deployed to capture the optical measurement.

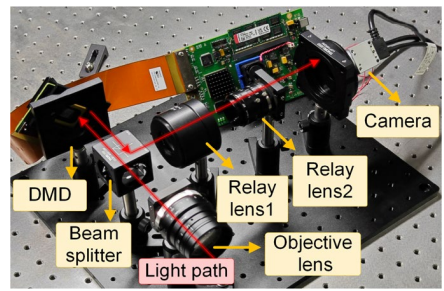

**Fig. 17.** Optical setup of CACTI system.

In the experimental systems, the DCR is set to 0.025 spv. Fig. 18 plots three frames of a domino video for visual comparison. Fig. 18(a) shows the results in AWGN channel transmission. When the SNR is 20 dB, the ICCI scheme recovers clearer spatial details than the SC schemes. When the SNR decreases to 0 dB, the ICCI scheme still achieves acceptable performance. However, the SC schemes fail to recover video information. Fig. 18(b) shows the results of slow fading channel transmission. When the SNR is 25 dB, the ICCI scheme achieves fewer artifacts than the SC schemes. When the SNR decreases to 5 dB, the content of the video can be recognized by applying the ICCI scheme. However, by adopting SC schemes, the video information is completely lost. The experimental results demonstrate the superiority of the ICCI framework in practical systems.

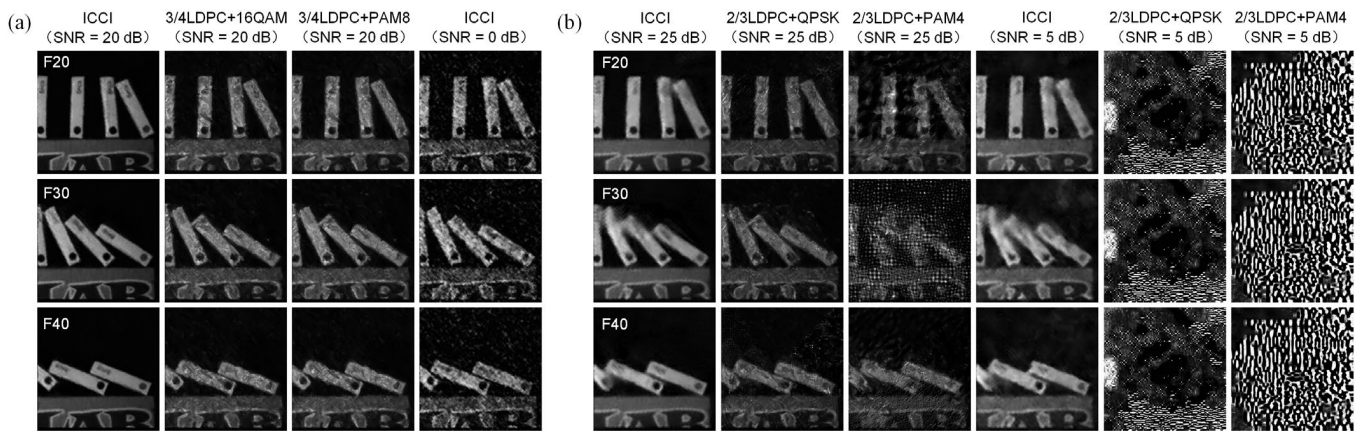

**Fig. 18.** Visual results obtained in (a) AWGN channel transmission and (b) slow fading channel transmission. The high-speed video information is acquired by the CACTI experimental system. FN denotes the N-th frame of the high-speed video.

## VII. Conclusion

We propose a novel ICCI framework that allows for the integration of communication and computational imaging to achieve end-to-end remote perception. By breaking the technical isolation between communication and computational imaging, this ICCI scheme achieves cross-domain integration. With the advantages of high throughput and low latency, this ICCI scheme can significantly reduce the bandwidth demand and overhead of terminal devices, thus providing low-cost and portable remote CV solutions for lightweight devices, such as mobile phones, intelligent surveillance cameras, and remote sensing drones. By leveraging the ICCI technology, remote perception nodes can be deployed in a more intelligent and miniaturized manner that facilitates the multifunctional integration of visual information sensing, transmitting, and computing. This advancement can improve the real-time vision perception and feedback capabilities of terminal devices to the physical world, subsequently enabling intelligent network access of visual terminals, such as virtual reality (VR), augmented reality (AR), and smart wearable devices. Therefore, it presents a viable solution for realizing the intelligent Internet of Everything (IoE). In addition, the ICCI framework allows terminal devices to realize multi-modal remote CV tasks by integrating different communications systems and CI systems. In Metaverse, this flexibility can improve the all-round perception abilities of machine terminals to the external world and enhance the interaction between human-machine-thing, which would contribute to a tighter connection between the physical and virtual spaces. Finally, the possibility of customizing the ICCI system for various downstream CV businesses can open a new era of goal-oriented services such as intelligent agriculture, intelligent factories, and intelligent transportation.